\documentclass[10pt,conference]{IEEEtran}
\usepackage{amsmath,amssymb,cite}
\begin{document}
\title{The P\'olya Urn: Limit Theorems, P\'olya Divergence, Maximum  Entropy and  Maximum Probability
}
\author{
\authorblockN{Marian Grendar}
\authorblockA{Department of Mathematics, FPV UMB \\
974 01 Banska Bystrica, Slovakia \\
Inst. of Mathematics and CS, Banska Bystrica, Slovakia \\
Inst. of Measurement Sci.,  Bratislava, Slovakia \\
marian.grendar@savba.sk} \and
\authorblockN{Robert K. Niven}
\authorblockA{
School of Aerospace, Civil and Mechanical Engineering \\
the University of New South Wales at ADFA \\
Northcott Drive \\
Canberra, ACT, 2600, Australia \\
r.niven@adfa.edu.au} }

\maketitle

\begin{abstract}
Sanov's Theorem and the Conditional Limit Theorem (CoLT) are
established for a multicolor P\'olya {Eggenberger} urn sampling
scheme, giving the P\'olya divergence  and the P\'olya extension to
the Maximum Relative Entropy (MaxEnt) method. P\'olya MaxEnt
includes the standard  MaxEnt as a special case. The universality of
standard MaxEnt  - advocated by an axiomatic approach to inference
for inverse problems - is challenged, in favor of a probabilistic
approach based on CoLT and the Maximum Probability principle.
\end{abstract}

\newtheorem{thm}{Theorem}
\newtheorem{ex}{Example}
\renewcommand{\QED}{\QEDopen}

\section{Introduction}

Consider an urn containing $\alpha_i>0$ balls of colors $i$,
$i=1,2,\dots, m$; let $m$ be finite. A single ball is drawn from the
urn, recorded and then returned together with $c\in\mathbb Z$ balls
of the same color. Assuming $-n\,c \le
\min(\alpha_1,\alpha_2,\dots,\alpha_m)$, the drawing is repeated $n$
times. This sampling is known as the multicolor P\'olya Eggenberger
(PE) urn scheme; c.f. \cite{EP}, \cite{Steyn}, \cite{JK}. Let
$\nu_i^n \triangleq \frac{n_i}{n}$ be the relative number of times a
ball of color $i$ is drawn in $n$ drawings. The vector $\nu^n
\triangleq [\nu^n_1,\nu^n_2,\dots,\nu^n_m]$ will be called type
\cite{CsiMT}, or $n$-type where necessary to stress that it is
induced by $n$ drawings.
Given the PE scheme, the probability $\pi(\nu^n; q, c)$ that
$n$-type $\nu^n$ will be drawn is (c.f. \cite{Steyn}, \cite{JK}):
\begin{equation}
\pi(\nu^n; q, c) \triangleq \frac{n!}{\prod_{i=1}^m n_i
!}\frac{\prod_{i=1}^m
\alpha_i(\alpha_i+c)\dotsm(\alpha_i+(n_i-1)c)}{N(N+c)\dotsm(N+(n-1)c)},
\label{eq_Polya_distrib}
\end{equation}
where $N\triangleq\sum_{i=1}^m \alpha_i$ and vector $q$ consists of
$q_i\triangleq\frac{\alpha_i}{N}$, $i=1,2,\dots,m$. The P\'olya
Eggenberger (PE) distribution (\ref{eq_Polya_distrib}) contains, as
prominent special cases, the multinomial distribution for $c=0$
(i.e., random sampling; identically and independently distributed
 ({\it iid}) outcomes), the multivariate hypergeometric distribution for $c=-1$ (i.e.,
sampling without replacement) and the multivariate negative
hypergeometric distribution for $c=1$; c.f. \cite{JK}.

Identify the set of possible colors (outcomes, states) with support
$\mathcal X$ of a random variable $X$. Following the notation of
\cite{CsiMT}, let $\mathcal{P}(\mathcal X)$ be the set of all
probability mass functions on $\mathcal X$. Let
$\mathcal{P}_n(\mathcal X)$ be the set of all possible $n$-types.
Finally, let $\Pi\subseteq\mathcal{P}(\mathcal X)$ {be the feasible
set of distributions} and $\Pi_n \triangleq
\Pi\cap\mathcal{P}_n(\mathcal X)$. The aim of this work is to
examine the Sanov Theorem for P\'olya sampling (i.e., the large
deviations behavior of $\pi(\nu^n\in\Pi; q, c)$), its associated
Conditional Limit Theorem (CoLT) and Gibbs Conditioning Principle
(GCP), and connections to the Maximum Probability (MaxProb)
principle \cite{Boltzmann}, \cite{Vincze72}, \cite{gg_what},
\cite{Niven1}, \cite{Niven2}, \cite{Niven3}. The asymptotic
investigations are conducted under the assumption that $N$,  $\beta
\triangleq \frac{N}{n}$ and $q$ may change with $n$  in such a way
that $q(n)\rightarrow q\in\mathcal{P}(\mathcal X)$ and
$\beta(n)\rightarrow\beta\in(0,1)$ as $n\rightarrow\infty$.

\section{P\'olya divergence}

Let $\beta\in (0,1)$, $c\neq 0$,  $p,q \in \mathcal{P}(\mathcal X)$,
and  $q + \beta c p \ge 0$. The P\'olya divergence $I(p\,||\,q;
\beta,c)$ of $p$ with respect to $q$ is given by:
\begin{multline*}
I(p\,||\,q;\beta,c) \triangleq I(p\,||\,q+\beta c p) +
\frac{1}{\beta c} I(q\,||\,q+\beta c p) \: + \\ + \frac{1+\beta
c}{\beta c}\log(1+\beta c),
\end{multline*}
where $I(a\,||\,b) \triangleq \sum_{i=1}^m a_i\log\frac{a_i}{b_i}$
is the Kullback Leibler (KL) divergence \cite{KL}, with standard
conventions. By the continuity argument, $I(p\,||\,q; \beta, 0)
\triangleq I(p\,||\,q)$.  When convenient, $I(p\,||\,q;\beta,c)$
will be replaced by  $I_\beta^c(p\,||\,q)$.

The following key properties of the P\'olya divergence are needed
for later analyses.

1) Non-negativity. $I_\beta^c(p\,||\,q) \ge 0$, with equality if and
only if (iff) $p=q$.

2) Lower semicontinuity. $I_\beta^c(p\,||\,q)$ is lower
semicontinuous in $p$, $q$. If $q_i>0$, for $i=1,2,\dots,m$, then
P\'olya divergence is continuous in the pair $p,q$.

3) Convexity in $p$, $q$. For any $\lambda\in[0,1]$ and $p,p',q,q'$,
it holds that  $\lambda I_\beta^c(p\,||\,q) +
(1-\lambda)I_\beta^c(p'\,||\,q') \ge I_\beta^c(\lambda p +
(1-\lambda)p'\,||\,\lambda q + (1-\lambda)q')$.

4) Partition inequality. If $\mathcal A \triangleq \{A_1, A_2,\dots,
A_k\}$ is a partition  of $\mathcal X$ and $p_\mathcal{A}(j)
\triangleq \sum_{i\in A_j} p_i$, $q_\mathcal{A}(j) \triangleq
\sum_{i\in A_j} q_i$, $j=1,2,\dots,k$, then $I_\beta^c(p\,||\,q) \ge
I_\beta^c(p_\mathcal{A}\,||\,q_\mathcal{A})$, with equality iff
$p(i|i\in A_j) = q(i|i\in A_j)$, $i\in A_j$, for each $j$.

5) Pinsker inequality. If $c\ge 0$, the total variation distance
$d(p,q)\triangleq \sum_{i=1}^m |p_i - q_i|$ is bounded as follows:
$I_\beta^c(p\,||\,q) \ge \frac{1}{2(1+\beta c)^2} d(p,q)$.

\begin{proof}
The properties can be established along standard lines (c.f.
\cite{CsiShi}, \cite{CT}, \cite{K}). In particular, Properties 1, 3
and 4 follow from the log-sum inequality. We outline the proof of
the Pinsker inequality for the P\'olya divergence. Since the
partition inequality holds for P\'olya divergence, from the standard
argument (c.f. \cite{CT}) it is sufficient to consider the binary
$\mathcal X$ with $\hat p \triangleq [p,1-p]$ and $\hat q\triangleq
[q,1-q]$, such that $p\ge q$, and find out under what restriction on
$\gamma$ the difference $g(q) \triangleq I_\beta^c(\hat p\,||\,\hat
q) - \gamma\, d^2(\hat p,\hat q)$ remains negative. Note that the
difference is $0$ for $p=q$, by Property 1. The first derivative
$g'(q)$ is $g'(q) \triangleq (q-p)\{\frac{1}{(q+\beta c p)(1+\beta c
- (q + \beta c p ))} + \frac{1}{\beta
c}\frac{1}{q-p}\log\frac{1+\beta c - (q+\beta c p)}{1-q} -
8\gamma\}$. Since by assumption $q\le p$, $q + \beta c p < p(1 +
\beta c)$. If $c\ge0$,  $y \triangleq 1+\beta c - (q+\beta c p)<
(1+\beta c)(1-q)$  and $y>1-q$. Hence, in order to assure negativity
of the derivative, $8\gamma\le \frac{4}{(1+\beta c)^2}$. Setting up
$\gamma=\frac{1}{2(1+\beta c)^2}$ establishes the bound.
\end{proof}

\section{Sanov Theorem for P\'olya Sampling}

Topological qualifiers are meant in topology induced on the
$m$-dimensional simplex by the usual topology on $\mathbb{R}^m$.
Following \cite{CsiMT}, for a set $\Pi\in \mathcal{P}(\mathcal X)$
and $q \in\mathcal{P}(\mathcal X)$, $\inf_{p\in\Pi}
I_\beta^c(p\,||\,q)$  is denoted by $I_\beta^c(\Pi\,||\,q)$.

\begin{thm}[P\'olya Sanov Thm.]
Let $\Pi$ be an open set. Let $q(n)\rightarrow q$,
$\beta(n)\rightarrow\beta \in (0,1)$, as $n\rightarrow\infty$. Then,
for $n\rightarrow\infty$,
$$
\frac{1}{n}\log\pi(\nu^n\in\Pi; q(n), c) = -I_\beta^c(\Pi\,||\,q).
$$
\end{thm}

\smallskip

\begin{proof}
The Method of Types \cite{CsiMT} approach to Large Deviations will
be used.

For $c=0$, the Sanov Theorem is already established, c.f.
\cite{Sanov}, \cite{CsiMT}, \cite{CsiST}. The rate function is just
the KL divergence, i.e., $I_\beta^0(\cdot\,||q)$.

The case of $c\neq 0$ will be divided into two subcases: $c>0$ and
$c<0$. The following inequalities are needed:

$i)$ $n\log n - n \le \log n! \le (n+1)\log n - n$, valid for $n>6$, 

$ii)$ $(b-1)\log b - (a-1)\log a - (b-a) < \log\Gamma(b)
-\log\Gamma(a) <  \left(b-\frac{1}{2}\right)\log b -
\left(a-\frac{1}{2}\right)\log a - (b-a)$, $0<a<b$; due to
\cite{KV}.

For $c\neq 0\wedge \frac{Nq_i}{c}\notin(\mathbb{Z}^{-})^m\wedge
\frac{N}{c}\notin\mathbb{Z}^{-}$, formula (1) can equivalently be
expressed as \cite{JK}:
\begin{equation}
\pi(\nu^n;q,c) =  \frac{n!}{\prod_{i=1}^m n_i
!}\frac{\Gamma\left(\frac{N}{c}\right)}{\Gamma\left(\frac{N}{c} +
n\right)}\prod_{i=1}^m\frac{\Gamma\left(\frac{Nq_i}{c} +
n_i\right)}{\Gamma\left(\frac{Nq_i}{c}\right)},
\end{equation}
where $\Gamma(\cdot)$ is the Gamma function.

Let $c>0$. Note then that the other restrictions under which (1) and
(2) are equivalent are not active, since $-nc \le \min(\alpha_1,
\alpha_2, \dots, \alpha_m)$. Applying the inequalities $i)$, $ii)$
to (2), $\frac{1}{n}\log\pi(\nu^n; q(n), c)$ is, for $n>6$,  bounded
from above by $U_n$ and from below by $L_n$:
\begin{gather*}
U_n = -I(\nu^n\,||\,q(n);\beta(n),c) + \frac{(m+1)\log n}{n} \: + \\ +\: \frac{1}{2n}(\log(1+\beta(n) c)  -\frac{1}{2n}\left(\sum_{i=1}^m\log\left(\frac{q_i(n)+\beta(n)c}{q_i(n)}\right)\right),\\
L_n = -I(\nu^n\,||\,q(n);\beta(n),c) \: + \\ + \:
\frac{1}{n}\left(\log(1+\beta(n) c)
-\sum_{i=1}^m\log\left(\frac{q_i(n)+\beta(n)
c}{q_i(n)}\right)\right).
\end{gather*}
To establish $L_n$ the standard ``trick'' of binding $\sum_{i=1}^m \log\nu^n_i$ from above by 
$-\sum_{i=1}^m\log m$ was used, in addition to $i)$ and $ii)$. The
non-divergence terms will be denoted $u_n$, $l_n$, respectively.

Let the cardinality of a set $\mathcal A$ be denoted $|\mathcal A|$.
$|\Pi| \le |\mathcal{P}_n(\mathcal X)| \le (n+1)^m$; c.f.\
\cite{CsiMT}. Thus,
\begin{gather*}
-I(\Pi_n\,||\,q(n);\beta(n),c) + l_n\le
\frac{1}{n}\log\pi(\nu^n\in\Pi; q(n), c) \le \\  \le
\frac{m\log(n+1)}{n}  -I(\Pi_n\,||\,q(n);\beta(n),c) + u_n.
\end{gather*}
Since $m$ is finite, all terms other than $I(\cdot)$ converge to
zero as $n\rightarrow\infty$. By assumption, $q(n)\rightarrow q$,
$\beta(n)\rightarrow\beta\in(0,1)$. Also, by assumption, $q_i(n)>0$,
for all $i$, thus $I(\cdot\,||\,\cdot;\beta(n),c)$ is continuous.
$\Pi$ is assumed to be open. Thus
$I(\Pi_n\,||\,q;\beta(n),c)\rightarrow I(\Pi\,||\,q;\beta,c)$ as
$n\rightarrow\infty$.

For $c\neq 0\wedge (1-\frac{Nq_i}{c})\notin(\mathbb{Z}^{-})^m\wedge
(1-\frac{N}{c})\notin\mathbb{Z}^{-}$, the formula (1) can
equivalently be expressed as:
\begin{equation}
\pi(\nu^n;q,c) =  \frac{n!}{\prod_{i=1}^m n_i !}\frac{\Gamma\left(1
-\frac{N}{c}- n\right)}{\Gamma\left(1 -
\frac{N}{c}\right)}\prod_{i=1}^m\frac{\Gamma\left(1
-\frac{Nq_i}{c}\right)}{\Gamma\left(1- \frac{Nq_i}{c} - n_i\right)}.
\end{equation}

Let $c<0$. Note then that the other restrictions under which (1) and
(3) are equivalent are not active, since $-nc \le \min(\alpha_1,
\alpha_2, \dots, \alpha_m)$. Applying the inequalities $i)$, $ii)$
to (3), the probability $\frac{1}{n}\log\pi(\nu^n; q(n), c)$ can be
bounded by $U_n$ from above and by $L_n$ from below, as $L_n = -
A(\nu^n\,||\,q(n);\beta(n),c)+ l_n$, $U_n = -
A(\nu^n\,||\,q(n);\beta(n),c) + u_n$, where
\begin{multline*}
A(\nu^n\,||\,q(n);\beta(n),c) \triangleq
\sum_{i=1}^m\nu_i^n\log\nu_i^n
+\frac{1+\beta(n)c}{\beta(n)c} \cdot \\
\cdot\log\left(-\frac{1+\beta(n)c}{\beta(n)c} + \frac{1}{n}\right)
- \frac{1}{\beta(n)c}\log\left(-\frac{1}{\beta(n)c} + \frac{1}{n}\right) \: + \\
+\: \sum_{i=1}^m\frac{q_i}{\beta(n)c}\log\left(-\frac{q_i}{\beta(n)c} + \frac{1}{n}\right) \: - \\
\:-\sum_{i=1}^m\frac{q_i+\beta(n)c\nu_i^n}{\beta(n)c}\log\left(-\frac{q_i+\beta(n)c\nu_i^n}{\beta(n)c}
+ \frac{1}{n}\right).
\end{multline*}
and $l_n$, $u_n$ stand for terms that converge to $0$ as
$n\rightarrow\infty$.

Using the same argument as for $c>0$,
$\frac{1}{n}\log\pi(\nu^n\in\Pi; q(n),c)$ is bounded
\begin{gather*}
-A(\Pi_n\,||\,q(n);\beta(n),c) + l_n\le
\frac{1}{n}\log\pi(\nu^n\in\Pi; q(n), c) \le \\ \le
\frac{m\log(n+1)}{n}  -A(\Pi_n\,||\,q(n);\beta(n),c) + u_n.
\end{gather*}
Since $m$ is finite, the terms other than $A(\cdot)$  converge to
zero, for $n\rightarrow\infty$. Since
$A(\cdot\,||\,q(n);\beta(n),c)$ is continuous, the argument used
above (case of $c>0$) implies that
$A(\Pi_n||\,q(n);\beta(n),c)\rightarrow I(\Pi\,||\,q;\beta,c)$, as
$n\rightarrow\infty$.
\end{proof}

\section{P\'olya Conditional Limit Theorem}

The P\'olya information projection $\hat{p}(\beta,c)$ (P\'olya
$I$-projection, or $I_\beta^c$-projection, for short) of $q$ on
$\Pi$ is defined as $\hat{p}(\beta,c)\triangleq\arg\inf_{p\in\Pi}
I_\beta^c(p\,||\,q)$. The standard $I$-projection \cite{CsiShi} is
the special $(c=0)$-case of the P\'olya $I$-projection.

The P\'olya Conditional Limit Theorem (CoLT) is an important
consequence of the P\'olya Sanov Theorem.

\begin{thm}[P\'olya CoLT]  Let $q(n)\rightarrow q$, $\beta(n)\rightarrow\beta \in (0,1)$, as $n\rightarrow\infty$. Let $\Pi$ be a convex, closed set.
Let $\hat{p}(\beta,c)$ be the $I_\beta^c$-projection of $q$ on
$\Pi$. Let $B(q,\epsilon)$ be the $\epsilon$-ball defined by the
total variation metric, centered at $q$. Then for any $\epsilon >
0$,
$$
\lim_{n\rightarrow\infty} \pi(\nu^n\in
B(\hat{p}(\beta,c),\epsilon)\,|\,\nu^n\in\Pi; q(n),c) = 1.
$$
\end{thm}

\smallskip

\begin{proof}
Let $B^C(\hat{p}(\beta,c),\epsilon)\triangleq\mathcal{P}(\mathcal
X)\backslash B(\hat{p}(\beta,c),\epsilon)$. Apply P\'olya Sanov
Theorem to $\pi(\nu^n\in B^C(\cdot)|\nu^n\in\Pi;
q(n),c)=\frac{\pi(\nu^n\in B^C(\cdot))}{\pi(\nu^n\in\Pi)}$. The
decay rate $I_\beta^c(B^C(\cdot)\,||\,q) - I_\beta^c(\Pi\,||\,q) >
0$. Since $\Pi$ is, by assumption, convex and closed, by convexity
of P\'olya information projection (Property 3) there is unique
$I_\beta^c$-projection of $q$ on $\Pi$. Types thus asymptotically
concentrate on it.
\end{proof}

P\'olya CoLT has the same interpretation as the standard,
$iid$-case, CoLT (see \cite{Vincze72}, \cite{Vasicek}, \cite{CC},
\cite{BS}, \cite{CsiME}): types induced by  PE sampling,
asymptotically conditionally (on the event $\nu^n\in\Pi$)
concentrate on the P\'olya information projection $\hat{p}(\beta,c)$
of $q$ on $\Pi$.

Setting $\Pi=\mathcal{P}(\mathcal{X})$ reduces P\'olya CoLT into its
special case: the Law of Large Numbers for PE sampling.

\section{Further Results}

By means of the P\'olya Sanov Theorem and the bounds used for its
proof, three additional results can be obtained.

\subsection{P\'olya Gibbs Conditioning Principle}

For the {\it iid} sampling there is a  claim, stronger than the Conditional Limit Theorem, known as Gibbs Conditioning Principle (GCP); c.f.\ \cite{CsiST,CsiMT,DZ}. Alongside of its 
proof \cite{CsiMT}, the following Gibbs Conditioning Principle for
PE sampling can be established.

\begin{thm}[P\'olya GCP]
Let $q(n)\rightarrow q$, $\beta(n)\rightarrow\beta \in (0,1)$, as
$n\rightarrow\infty$. Let $\Pi$ be a convex, closed set. Let
$\hat{p}(\beta,c)$ be the $I_\beta^c$-projection of $q$ on $\Pi$.
Then for a fixed $t$,
\begin{multline*}
\lim_{n\rightarrow\infty}\pi(X_1=x_1,\dots,X_t=x_t\,|\nu^n\in\Pi;
q(n),c) = \\ = \prod_{l=1}^t\hat{p}_l(\beta,c).
\end{multline*}
\end{thm}

Loosely put, asymptotically, conditionally upon the event
$\nu^n\in\Pi$, a fixed-length sequence of drawn colors behaves as if
it was identically and independently drawn from the P\'olya
information projection $\hat{p}(\beta,c)$ of $q$ on $\Pi$.

Its $t=1$ special case can be established for $c=0$ by means of the
Pythagoras property of the $I$-projection and the Pinsker
inequality; see \cite{CT}. This approach does not carry on to $c\neq
0$, as the Pythagoras property does not hold for
$I_\beta^c$-projection with $c\neq 0$.

\subsection{\label{sect_MaxProb}Maximum Probability - Maximum Entropy Correspondence}

Let $\hat\nu^n(\beta,c)\triangleq\arg\sup_{\nu^n\in\Pi_n}\pi(\nu^n;
q, c)$ be the P\'olya $\mu$-projection ($\mu_\beta^c$-projection,
for short) of $q$ on $\Pi_n$; i.e., the supremum-probable $n$-type
in $\Pi_n$.  Using the $U_n$, $L_n$ bounds (c.f.\ proof of P\'olya
Sanov Theorem), the asymptotic identity of P\'olya $\mu$-projections
and P\'olya $I$-projections, can be established along the lines of
\cite{g_cet}.
\begin{thm}[MaxProb/MaxEnt]
Let $q(n)\rightarrow q$, $\beta(n)\rightarrow\beta \in (0,1)$, as
$n\rightarrow\infty$. Let $\mathcal{M}_n(\beta(n),c)$ be a set of
all $\mu_{\beta(n)}^c$-projections of $q(n)$ on $\Pi_n$. Let
$\mathcal{I}_\beta^c$ be a set of all $I_\beta^c$-projections of $q$
on $\Pi$. Then, for $n\rightarrow\infty$, $\mathcal{M}_n(\beta(n),c)
= \mathcal{I}(\beta,c)$.
\end{thm}

This permits a deeper interpretation of the  P\'olya CoLT.
Informally: types, asymptotically conditionally (upon $\nu^n\in\Pi$)
concentrate on the most probable type. Even more loosely put, the
most probable is asymptotically conditionally the only possible.

The asymptotic identity of P\'olya $\mu$-projections and P\'olya
$I$-projections is illustrated by the following Example.

\begin{ex}
Let $\mathcal X =  \{1, 2, 3, 4\}$, i.e., there are four colors,
associated with the numbers. Let $\Pi = \{p: \sum_{i=1}^4 p_i x_i =
3.2, \sum_{i=1}^m p_1 = 1\}$. Let $n = 10, 50, 100, 1000$ and $N(n)
= 100, 500, 1000, 10000$, so that $\beta_n = 0.1$, for all
considered $n$. Let $q(n) = q = [21,\  25,\  31,\  23]/100$ for all
considered $n$. For each $n$, let $c \in \{-2, 1, 0, 1, 5, 10\}$.
The Table in Appendix A contains the P\'olya $\mu$-projection
$\hat{\nu}^n$  of $q(n)$ on $\Pi_n$. In the last block of the Table,
the P\'olya $I$-projection $\hat p$ of $q$ on $\Pi$ is presented,
for each considered $c$.
\end{ex}

\subsection{P\'olya Conditional Equi-concentration of Types}

The Conditional Equi-concentration of Types (CET) on
$I$-projections, an extension of CoLT to the case of $\Pi$ admitting
more than one $I$-projection, is discussed at \cite{g_cet}.
Similarly, CET holds also for P\'olya $I$-projections.

\section{Applications and implications}

P\'olya CoLT has similar applications (and implications) as the
standard $iid$ CoLT (see \cite{CsiME,g_cet}), but holds in a broader
context of  PE sampling, which encapsulates the $iid$ one. We will
briefly discuss two applications.

\subsection{P\'olya MaxEnt}

The Boltzmann P\'olya Inverse Problem (BPIP) contains the Boltzmann
Jaynes Inverse Problem \cite{g_cet} as its special, {\it iid}, case.
BPIP is constituted by  the information-pentad $\{\mathcal X, Nq, n,
c, \Pi\}$ under which the objective is to select a type (one or
more) from $\Pi$. Three examples of BPIP are below.

\begin{ex}
A  network containing $i=1,2,\dots, m,$ critical nodes, each with
$\alpha_i$ branches, is  accessed  by users. Each time a connection
to node $i$ is accessed, $c$ new connections to this node are
established ($c=-1$ indicates simple congestion, and $c<-1$,
accelerated congestion). Assuming that  the number of transactions
$n$ is known, the objective is to select a type of connections to
the nodes from $\Pi\equiv\mathcal{P}_n(\mathcal X)$.
\end{ex}

\begin{ex}
A stock exchange is established with $N$ equiprobable shares, with
$\alpha_i$ of share $i$. After a trade in stock $i$, $c$ new shares
are issued in it ($c<0$ indicates withdrawal of shares). The
transactions are constrained by the mean value of trades in a given
period. Given the feasible set $\Pi_n$ of $n$-transactions
determined by the mean value of trades, and the other above
described information, the objective is to select an $n$-type of
transactions.
\end{ex}

\begin{ex}
Let $n$ out of $N$ quantum mechanics particles be distributed among
$m$ energy levels according to {the PE}  sampling scheme with
initial distribution $q$ and parameter $c$. Let instead of the
actual energy distribution ($n$-type) only the mean value of energy
of $n$ particles be available. Given this information, the objective
is to select an $n$-type from the feasible set.
\end{ex}

BPIP is under-determined and in this sense {is an} ill-posed inverse
problem. The indeterminacy of the problem translates into {a}
multitude of possible methods for its solution. From an infinite set
of possible methods of solving BPIP, such a method has to be
selected that, for $n\rightarrow\infty$, does not violate P\'olya
CoLT. Clearly, selection of the P\'olya $I$-projection of $q$ on
$\Pi_n$ satisfies the above requirement of asymptotic consistency.
This selection scheme could reasonably be called the P\'olya Maximum
Relative Entropy (MaxEnt) method, where the P\'olya relative entropy
is defined as the negative of the P\'olya divergence;
$H(p\,||\,q;\beta,c) \triangleq - I(p\,||\,q;\beta,c)$. Note that
from the point of view of maximization over $p$, the P\'olya
relative entropy effectively reduces to:
\begin{displaymath}
 - \sum p_i\log p_i + \sum\left(p_i + \frac{q_i}{\beta c}\right)\log(q_i + \beta c p_i).
\end{displaymath}

The other way for solving/regularizing BPIP that is asymptotically
consistent is Maximum Probability (MaxProb), that selects the
$\mu_\beta^c$-projection of $q$ on $\Pi_n$. By P\'olya
MaxProb/MaxEnt the two methods asymptotically coincide, but for
finite $n$ they make, in general, a different choice.

The feasible set $\Pi$ can for instance (as in the above Examples 3
and 4) be formed by moment-consistency constraints $\Pi=\{p:
\sum_{i=1}^m p_i u_j(x_i) = a_j, j = 0, 1,2,\dots, J\}$, where
$u_j(\cdot)$ is a given real-valued function, $u_0(\cdot) \triangleq
1$; $a_j$ is given number, $a_0 \triangleq 1$; such a feasible set
is also known as the linear family of distributions. The P\'olya
$I$-projection of $q$ on the linear family of distributions $\Pi$ is
 then implicitly given by:
\begin{equation}
\hat{p}_i(\beta,c) = \frac{q_i e^{-\sum_{j=0}^J \lambda_j
u_j(x_i)}}{1-\beta ce^{-\sum_{j=0}^J \lambda_j u_j(x_i)}}.
\end{equation}

\subsubsection{Distribution of anyons}
Not surprisingly, for $c=0$, the probability distribution (4) turns
into the familiar exponential (Maxwell-Boltzmann) form of the
$I$-projection on the linear family; \cite{CsiShi}. For $c=-1$, the
distribution gives the Fermi Dirac distribution whilst for $c=+1$,
the distribution gives the Bose Einstein distribution. These are
generalizations of the standard Bose Einstein and Fermi Dirac
distributions, in the sense that a general (not necessarily uniform)
sampling distribution $q$ is assumed \cite{NG}. In this respect it
is worth recalling the Example 4 and  noting that the PE
distribution (4) contains an ansatz distribution of
quantum-mechanical  anyons  (i.e., particles with properties
intermediate between those of bosons and fermions; \cite{W})
proposed at \cite{ANS} as its special (uniform $q$) case. This in
our view, provides both a probabilistic underpinning of the Acharya
\& Narayana Swamy \cite{ANS} distribution of anyons as well as its
extension to the non-uniform sampling case. Further discussion will
be given elsewhere \cite{NG}.

\subsubsection{Limitation of axiomatic approach to linear inverse problems}

We would like to stress that for $c\neq 0$ the standard  MaxEnt
method \cite{Jaynes}, \cite{Kullback} (i.e., selection of
$I$-projection of $q$ on $\Pi_n$), when applied to BPIP, does not
satisfy the requirement of asymptotic consistency. Thus, although
the standard MaxEnt is advocated by an axiomatic approach as the
logically consistent way of solving ill-posed inverse problems with
$\Pi$ defined by the moment-consistency constraints (c.f.
\cite{CsiWhy}, \cite{CsiME}), the method, when applied under PE
sampling with $c\neq 0$, violates the P\'olya Conditional Limit
Theorem. This reveals a limitation of {the} axiomatic approach to
inference in the inverse problems context.

\subsection{Rare events simulation}

P\'olya CoLT and P\'olya GCP can be used for rare events simulation
in {the} context of PE sampling, in the same way that the standard
CoLT and GCP are used in the {\it iid} sampling; c.f.\
\cite{Bucklew}.

\section{Summary}

The standard Conditional Limit Theorem (CoLT) \cite{CT} for {\it
iid} sampling provides a probabilistic justification (c.f.
\cite{Vincze72}, \cite{CsiME}) of  MaxEnt method in the context of
so-called Boltzmann Jaynes Inverse Problem (BJIP), \cite{g_cet}. In
\cite{gg_what} it was suggested that MaxEnt can be viewed as an
asymptotic instance of the MaxProb method, {which} under the limited
information {available to the} BJIP, selects the type (i.e., the
empirical distribution) with the highest probability of occurrence,
from {the} given data-sampling distribution. It was proposed in
\cite{Vincze72}, \cite{Niven2}, \cite{Niven3}, that MaxProb can be
considered in a broader context; in particular under sampling
schemes other than the random (i.e., {\it iid}) sampling. There it
was also pointed out that every sampling scheme might be associated
{with} its own instance of MaxProb and its own {relative} entropy
maximization method. For a particular sampling scheme (or,
probabilistic question of certain form, in general) and adjoint
inverse problem, the relevant entropy maximization can be discovered
by considering the associated CoLT. The relevant CoLT, in turn,
provides probabilistic justification of the associated relative
entropy maximization method in the context of the inverse problem.
Motivated by these observations, in this work we have established
CoLT for {the} PE sampling scheme and discussed some of its
consequences and applications.

\section{Notes on literature}

An early physics-motivated work that extends Boltzmann's Maximum
Probability principle, steps into  the direction of Sanov Theorem
for non-{\it iid} sampling and contains a few views ahead of its
time is Vincze's \cite{Vincze72}; see also \cite{Vincze}.  For
sampling without replacement (i.e., $(c=-1)$-case of PE sampling),
{the} Sanov Theorem was established by \cite{DZ}. {A} communications
channel with P\'olya noise has been considered at \cite{AF}.

\section*{Acknowledgment}

One of the authors (M.G.) wishes to acknowledge hospitality of the
School of Computer Science and Engineering of the University of New
South Wales (UNSW), Sydney, as well as of the UNSW at ADFA,
Canberra, where this study was conducted. Supported by VEGA grant
1/3016/06 and Australian Research Council grant no. DP0210999.
Special thanks to Arthur Ramer.

\section{Appendix A}

The Table 1  illustrates Maximum Probability to P\'olya Maximum
Entropy convergence, for PE sampling; c.f.\ Sect. V.B.

\begin{table}
\renewcommand{\arraystretch}{1.3}
\caption{MaxProb to P\'olya MaxEnt convergence} \centering
\begin{tabular}{ | l | l  l  l  l |}
  \hline
  & & $\hat\nu^n$ & &  \\
  \hline
  n=10 &  &  &  & \\
  \hline
  c=-2 & 0 & 0.2 & 0.4 & 0.4  \\
  c=-1 & 0 & 0.2 & 0.4 & 0.4  \\
  c=0  & 0 & 0.2 & 0.4 & 0.4  \\
  c=1  & 0.1 & 0.1 & 0.3 & 0.5  \\
  c=5  & 0.1 & 0.1 & 0.3 & 0.5  \\
  c=10 & 0 & 0.2 & 0.4 & 0.4  \\
  \hline
  n=50 &  &  &  &   \\
  \hline
  c=-2 & 0.06 & 0.14 &  0.34 & 0.46  \\
  c=-1 & 0.06 & 0.14 &  0.34 & 0.46  \\
  c=0  & 0.06 & 0.14 &  0.34 & 0.46  \\
  c=1  & 0.06 & 0.14 &  0.34 & 0.46  \\
  c=5  & 0.06 & 0.14 &  0.34 & 0.46  \\
  c=10 & 0.06 & 0.14 &  0.34 & 0.46  \\
  \hline
  n=100 &  &  &  & \\
  \hline
  c=-2  & 0.05 & 0.15 &  0.35 &  0.45  \\
  c=-1  & 0.06 & 0.14 &  0.34 &  0.46   \\
  c=0   & 0.06 & 0.14 &  0.34 &  0.46   \\
  c=1   & 0.06 & 0.14 &  0.34 &  0.46   \\
  c=5   & 0.07 & 0.14 &  0.31 &  0.48   \\
  c=10  & 0.07 & 0.14 &  0.31 &  0.48   \\
  \hline
  n=1000 &  &  &  & \\
  \hline
  c=-2  & 0.056 &  0.142 & 0.348 & 0.454   \\
  c=-1  & 0.060  & 0.141  & 0.338  & 0.461   \\
  c=0 &  0.062 & 0.141 & 0.332 & 0.465    \\
  c=1 &  0.064 & 0.141 &  0.326 & 0.469   \\
  c=5 &  0.070 & 0.140  & 0.310  & 0.480    \\
  c=10 & 0.073 & 0.140 &  0.301 & 0.486 \\
  \hline
  & & $\hat p$ &  &  \\
  \hline
  c=-2  &  0.05628 & 0.14179  & 0.34759  & 0.45434  \\
  c=-1  &  0.05974 & 0.14126  & 0.33826  & 0.46074   \\
  c=0   &  0.06241 & 0.14085  & 0.33108  & 0.46566   \\
  c=1   &  0.06453 & 0.14052  & 0.32537  & 0.46958   \\
  c=5   &  0.06997 & 0.13968  & 0.31072  & 0.47962  \\
  c=10  &  0.07357 & 0.13913  & 0.30102  & 0.48628  \\
  \hline
\end{tabular}
\end{table}

\end{document}